\documentclass[twocolumn]{aastex631}

\usepackage{xcolor}
\usepackage{amsmath}
\usepackage{hyperref}
\usepackage{xspace}
\usepackage{natbib}
\usepackage{graphicx}
\graphicspath{{./figures/}}

\newcommand{\Lya}{Ly$\alpha$}

\newcommand{\simba}{{\sc Simba}\xspace}

\defcitealias{Danforth:2016}{D16}
\def\DanforthCOS/{\citetalias{Danforth:2016}}

\shorttitle{Long-range AGN feedback affects the low$-z$ Lyman-$\alpha$ forest}
\shortauthors{Tillman et al.}

\begin{document}

\title{Efficient long-range AGN feedback affects the low redshift Lyman-$\alpha$ forest}

\author[0000-0002-1185-4111]{Megan Taylor Tillman}
\affiliation{Department of Physics and Astronomy Rutgers University,  136 Frelinghuysen Rd, Piscataway, NJ 08854, USA}

\author[0000-0001-5817-5944]{Blakesley Burkhart}
\affiliation{Department of Physics and Astronomy, Rutgers University,  136 Frelinghuysen Rd, Piscataway, NJ 08854, USA}
\affiliation{Center for Computational Astrophysics, Flatiron Institute, 162 Fifth Avenue, New York, NY 10010, USA}

\author[0000-0002-8710-9206]{Stephanie Tonnesen}
\affiliation{Center for Computational Astrophysics, Flatiron Institute, 162 Fifth Avenue, New York, NY 10010, USA}

\author[0000-0001-5803-5490]{Simeon Bird}
\affiliation{University of California, Riverside, 92507 CA, U.S.A.}

\author[0000-0003-2630-9228]{Greg L. Bryan}
\affiliation{Center for Computational Astrophysics, Flatiron Institute, 162 Fifth Avenue, New York, NY 10010, USA}
\affiliation{Department of Astronomy, Columbia University, 550 W 120th Street, New York, NY 10027, USA}

\author[0000-0001-5769-4945]{Daniel Angl\'es-Alc\'azar} 
\affiliation{Department of Physics, University of Connecticut, 196 Auditorium Road, U-3046, Storrs, CT 06269-3046, USA}
\affiliation{Center for Computational Astrophysics, Flatiron Institute, 162 5th Ave, New York, NY 10010, USA}

\author[0000-0003-2842-9434]{Romeel Dav\'e}
\affiliation{Institute for Astronomy, University of Edinburgh, Royal Observatory, Edinburgh, EH9 3HJ, UK}
\affiliation{University of the Western Cape, Bellville, Cape Town 7535, South Africa}

\author[0000-0002-3185-1540]{Shy Genel}
\affiliation{Center for Computational Astrophysics, Flatiron Institute, 162 Fifth Avenue, New York, NY 10010, USA}
\affiliation{Columbia Astrophysics Laboratory, Columbia University, 550 West 120th Street, New York, NY 10027, USA}

\begin{abstract}

Active galactic nuclei (AGN) feedback models are generally calibrated to reproduce galaxy observables such as the stellar mass function and the bimodality in galaxy colors.
We use variations of the AGN feedback implementations in the IllustrisTNG (TNG) and \simba cosmological hydrodynamic simulations to show that the low redshift Lyman-$\alpha$ forest can provide constraints on the impact of AGN feedback. 
We show that TNG over-predicts the number density of absorbers at column densities $N_{\rm HI}
< 10^{14}$ cm$^{-2}$ compared to data from the Cosmic Origins Spectrograph (in agreement with previous work), and we demonstrate explicitly that its kinetic feedback mode, which is primarily responsible for galaxy quenching, has a negligible impact on the column density distribution (CDD) of absorbers. 
In contrast, we show that the fiducial \simba model which includes AGN jet feedback is the preferred fit to the observed CDD of the $z = 0.1$ Lyman-$\alpha$ forest across five orders of magnitude in column density.
We show that the \simba results with jets produce a quantitatively better fit to the observational data than the \simba results without jets, even when the UVB is left as a free parameter.
AGN jets in \simba are high speed, collimated, weakly-interacting with the interstellar medium (via brief hydrodynamic decoupling) and heated to the halo virial temperature. 
Collectively these properties result in stronger long-range impacts on the IGM when compared to TNG's kinetic feedback mode, which drives isotropic winds with lower velocities at the galactic radius.
Our results suggest that the low redshift Lyman-$\alpha$ forest provides plausible evidence for long-range AGN jet feedback.
\end{abstract}


\section{Introduction}

    The \Lya~forest is a series of absorption line features that originate from the distribution of intergalactic gas along the line of sight to a background source. 
    These absorption lines provide a slew of statistical measurements including the flux PDF and power spectrum, column density distribution (CDD), and the widths of the absorption lines themselves (often referred to as b-values).
    This makes the \Lya~forest a powerful diagnostic tool at high redshifts $z\gtrsim2$ for the properties of dark matter (DM) and the thermal state of the intergalactic medium (IGM) \citep{Gunn:1965,Palanque-Delabrouille:2013,Viel:2013, Puchwein:2018,bolton:2021}. The forest also provides a census of the temperature and matter density at different epochs in the IGM \citep{Altay:2011,Rahmati:2013,Hernquist:1996,Hiss:2018,Walther:2019,Chabanier:2020}. 

    The absorbing gas that creates the forest is often assumed to be in photoionization equilibrium with the ionizing background radiation.
    This assumption gives rise to a \Lya~optical depth of $\tau=-\rm{ln ~F} \propto N_{\rm HI}^2 T^{-0.7} \Gamma_{\rm HI}^{-1}$, where F is the normalized transmitted flux, $\rm{N_{\rm HI}}$ is the hydrogen number density, T is the IGM gas temperature, and $\Gamma_{\rm HI}$ is the hydrogen photoionization rate. 
    While it is typical to assume photoionization equilibrium, simulations differ in whether they assume thermal equilibrium.
    Thus, if the temperature is fully set by photoelectric heating by the integrated ultraviolet (UV) emission from background stars and quasars,  one may relate the statistics of the forest to the ultraviolet background (UVB) as well as to the underlying dark matter distribution.
    Indeed, the assumption of photoionization equilibrium in the IGM provides an excellent match to the observed properties of the \Lya~forest at redshift $z \geq 2$ \citep{Katz:1996,Hernquist:1996} and  allows for a quantitative connection between the \Lya~ forest and the dark matter density field \citep{Weinberg:1999, Peeples:2010}.

    Despite the vast utility of the $z \geq 2$ \Lya~forest in constraining our conceptual understanding of the IGM, it has been difficult to reconcile simulations/models of the low redshift \Lya~forest with observations.  
    Observing the \Lya~forest at $z \leq2$ requires state-of-the-art spectrographs that are above Earth's atmosphere and cover the rest-frame far-ultraviolet (FUV) band \citep{Danforth:2016,Gurvich2017,Viel:2017,Khaire:2019,Christiansen:2020}.
    Observations with the Cosmic Origins Spectrograph (COS) aboard the Hubble Space Telescope (HST), coupled with previous space missions in the UV such as FUSE \citep{Danforth:2005}, have enabled studies of the \textit{low-redshift} \Lya~forest in great statistical detail and with high sensitivity \citep{Tripp:2008,Meiring:2011,Khaire:2019,Danforth:2016,Kim:2020}. 
    In particular, \citet{Danforth:2016} (henceforth D16) produced an extensive catalog of low redshift absorbers by building on previous HST catalogs. 
    This catalog has enabled the study of the low redshift forest in great statistical detail as it contains over 5000 absorbers and probes column densities as low as $N_{\rm HI} \approx 10^{12.5} \rm{cm}^{-2}$. 
    Confronted with the D16 dataset, cosmological hydrodynamical simulations have struggled to reproduce the observed statistics of the low redshift \Lya~forest \citep{Kollmeier:2014}, and require more ionizing UVB models than \citet{Haardt:2012} \citep[e.g.][]{Gaikwad:2017, KhaireUVB:2019, Puchwein:2019, FG:2020} and/or additional non-standard, heating sources \citep[][focusing on large-scale environment and galactic/AGN feedback]{Viel:2017, Gurvich2017, Tonnesen:2017, Christiansen:2020}. 
    
    \citet{Gurvich2017} was the first to point out that updated AGN feedback models were likely required to fully resolve the low redshift \Lya~forest discrepancy between simulations and observations. They point out that AGN could provide additional heating beyond the standard UVB photoionization equilibrium models. 
    Later studies confirmed that AGN feedback can have a dramatic heating effect on the IGM, especially at lower redshifts \citep{Viel:2017, Christiansen:2020, Burkhart_2022}.
    In particular, \citet{Christiansen:2020} explore variations on the \simba simulation that utilize different implementations of the AGN feedback.
    They explore the properties of the IGM, focusing on the \Lya~forest mean flux decrement, and find that the jet mode in particular is vital in reproducing what is observed at low redshift.
    In a similar vein, \citet{Burkhart_2022} found that the different AGN feedback models in the Illustris and IllustrisTNG (henceforth TNG) cosmological suites produced very different \Lya~forest statistics (i.e. b-distribution, CDD, flux power spectrum), despite having the same UVB model.
    However the exploration of the entire column density range in the context of the \simba AGN jet feedback has not been conducted, and a comparison between the \simba and TNG CDDs would help disentangle the effects of different AGN feedback modes.
    
    By $z=0.1$ it has been shown that AGN feedback models can remake the entire thermal state of the IGM, including altering the hot gas and the neutral fraction \citep{Martizzi:2019}. 
    Thus the nature of the AGN model can be constrained not only by matching galaxy properties but also via observations of the circumgalactic medium (CGM), intracluster medium (ICM), and IGM. 
    In this work we focus directly on the CDD of the \simba cosmological suite used in \citet{Christiansen:2020} to further investigate the proposed match of the \Lya~forest produced with AGN jet feedback, and the \citet{Haardt:2012} UVB to the D16 dataset. 
    We then compare the \simba AGN feedback model to that of TNG and discuss differences that might cause the variation seen in the two simulations' CDDs.
   
    The paper is organized as follows: in Section~\ref{sec:methods} we describe the simulations we analyze with a focus on the AGN feedback models, how the \Lya~forest spectra is generated, and how the CDD is calculated.
    In Section \ref{sec:res} we present the resulting $z=0.1$ CDDs from \simba simulations that include different AGN feedback modes. We conduct a comparison to the TNG CDD \citep[recently analyzed in][]{Burkhart_2022} and we explore an additional TNG run that removes the kinetic AGN feedback mode.
    In Section~\ref{sec:dis} we discuss what the results from these comparisons reveal about the AGN jet feedback in \simba, and we discuss differences between the TNG and \simba models that motivate the use of AGN jet feedback. We conduct a post-processing UVB correction to the \simba results to explore any degeneracy from the effects of the AGN jets vs.\ the UVB model on the CDD. Finally, we summarize our findings and discuss the necessary next steps in Section~\ref{sec:con}.

    \begin{figure*}
        \centering
        \includegraphics[width=0.7\linewidth,trim={6in 0 0 0},clip]{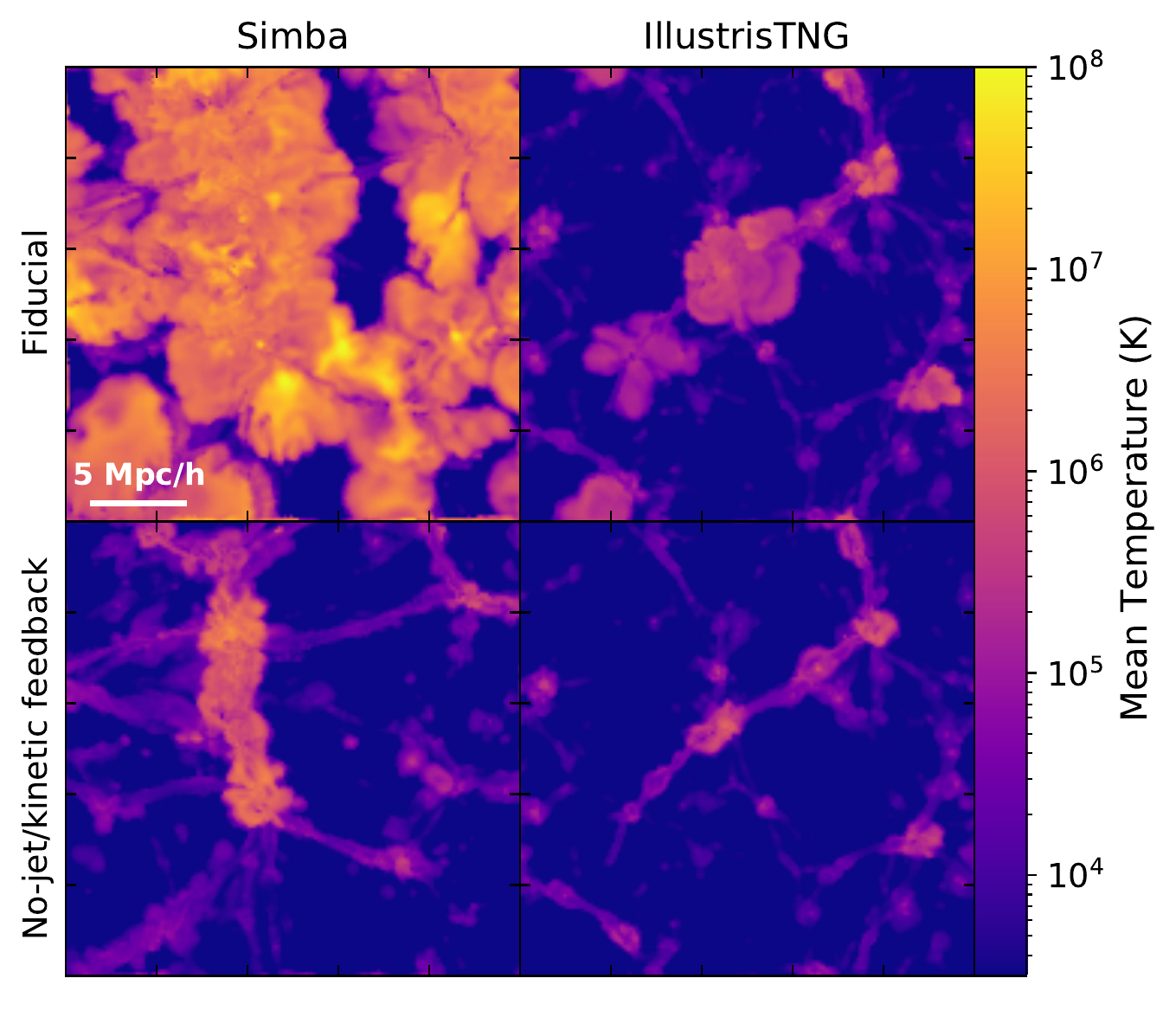}
        \caption{The $z=0.1$ mass weighted mean temperature projections for the \simba feedback variations and the CAMELS TNG runs. Each projection is over a depth of $525$ kpc/h (corresponding to the absorber length discussed in section \ref{sec:gen_CDDF}). The top rows display the fiducial runs and the bottom rows display the runs without AGN jet/kinetic feedback for \simba (left) and TNG (right). The \simba projection is a 25 by 25 Mpc/h portion of the full (50 by 50 Mpc/h) box while the CAMELS TNG projection is the full 25 by 25 Mpc/h box.}
        \label{fig:Tprojs} 
    \end{figure*}

    \begin{figure*}
        \centering
        \includegraphics[width=0.7\linewidth,trim={0 0 6in 0},clip]{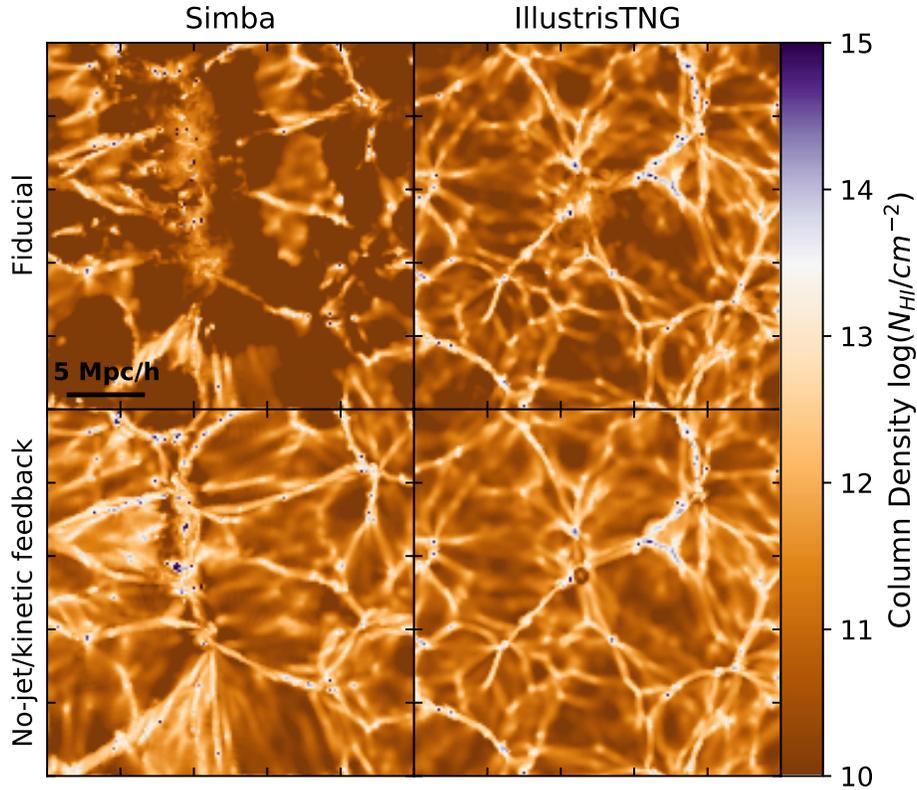}
        \caption{The same as Figure \ref{fig:Tprojs} but instead of a temperature projection, a \Lya~column density projection over a single 525 kpc/h slice is shown. These projections show how rare column densities of $N_{\rm HI} \gtrsim 10^{15}$ cm$^{-2}$ are.}
        \label{fig:CDprojs} 
    \end{figure*}

\section{Numerical Simulations} \label{sec:methods}

In this study we explore results from the \simba and IllustrisTNG (TNG) simulations.
In this section we briefly describe these simulations and the AGN feedback models they implement. 
The flagship Simba and IllustrisTNG100-1 simulations are run in a 100 and 75 Mpc/h box respectively but in this study we explore ``small box'' runs that vary the AGN feedback physics in the simulations. 
This allows us to analyze the effects of the different AGN feedback models on the \Lya~forest. 
We discuss more on these small box runs as compared to the larger box simulation runs throughout this section.
We discuss the resolution and box size of the simulations explored in this study and their effects on the CDD in the Appendix.
In short, we find the CDD of the small box runs to be converged with the CDD of the larger box simulations.

    \subsection{\simba}
        \simba is the next generation of the {\sc Mufasa}\xspace cosmological galaxy formation simulations \citep{dave:2016} run with GIZMO’s meshless finite mass hydrodynamics \citep{hopkins:2015}, and employs a number of state of the art subgrid physical processes to form realistic galaxies. 
        The GIZMO gravity solver is based on GADGET-3 \citep{Springel:2005} and evolves dark matter and gas together including gravity and pressure forces and follows shocks via a Riemann solver with no artificial viscosity.
        The \simba simulations use the following cosmological parameters: $\Omega_m = 0.3$, $\Omega_\Lambda$ = 0.7, $\Omega_b = 0.048$, $H_0 = 68$ km/s/Mpc, $\sigma_8 = 0.82$, and $n_s = 0.97$.
        We briefly review the essential aspects of the code here with a focus on the AGN model.  
        For more details on these implementations we refer the reader to  \citet{dave:2019}.
        
        Star formation is based on a Kennicutt–Schmidt Law \citep{Kennicutt:1998}  scaled by the H$_2$ fraction,
        which is calculated for each particle using its local column density and metallicity following \citet{krumholz:2011}.
        Galactic outflows are implemented as kinetic decoupled two-phase winds with an updated mass-loading factor based on particle tracking results from the Feedback in Realistic Environments (FIRE) zoom-in simulations \citep{DAA:2017b}.  
        The production of 11 elements (H, He, C, N, O, Ne, Mg, Si, S, Ca, and Fe) are tracked from Type II and Ia supernovae and from stellar evolution.
        Relevant for the IGM, photoionization heating and radiative cooling are implemented using GRACKLE-3.1 \citep{Grackle:2017} assuming ionization but not thermal equilibrium.  
        GRACKLE assumes a \citet{Haardt:2012}  ionizing background modified to account for self-shielding \citep{Rahmati:2013}.

    \subsubsection{The Fiducial \simba AGN Feedback Model}

        In the \simba simulations AGN feedback is modeled in two main discrete modes: a radiative mode at high Eddington ratios and a jet mode at low Eddington ratios \citep[both implemented as kinetic outflows;][]{DAA:2017a}, with the latter accompanied by X-ray photon energy feedback \citep{Choi:2012}. 
        
        The radiative mode drives multi-phase gas in winds at velocities of $\sim 500-1500$ km/s. 
        The transition to jet mode feedback occurs for SMBHs with Eddington ratios of $\eta$ $< 0.2$ and masses of $M_{\rm BH} \geq 10^{7.5} M_\odot$. Full jet velocity is reached when $\eta<0.02$.
        The gas ejected in these jets has a velocity that increases with lower $\eta$ and higher $M_{\rm BH}$, with a cap at 7000 km/s. 
        This results in maximum wind speeds in the jet feedback mode of $\sim 8000$ km/s when the appropriate criterion are met.
        
        The gas ejected in jets is decoupled from the hydrodynamics and cooling for a length of time scaling with the Hubble time at the moment of ejection (this results in a decoupling of $\sim 0.5$ Myr at $z=2.0$ to $\sim 1.5$ Myr at $z=0$). 
        As a result, the AGN jets can travel distances up to $\sim 10$ kpc before their energy begins to be deposited. 
        The temperature of the expelled gas in these jets is raised to the virial temperature of the halo ($T_{\rm vir} = 9.52 \times 10^7 (M_{\rm halo}/10^{15} M_\odot)^{1/3}$ K). 
        The gas is also ejected in a purely bipolar highly collimated fashion, along the angular momentum vector of the inner galactic disk.
        
        X-ray feedback occurs in conjunction with jet feedback but only when the maximum 7000 km/s velocity increase occurs and the galaxy has a gas to stellar mass ratio of $M_\textnormal{gas}/M_* < 0.2$. 
        The X-ray feedback heats the non-ISM gas surrounding the SMBH accretion kernel. 
        This heating decreases with distance to the black hole. 
        For ISM gas, half of the X-ray energy is applied as kinetic energy in a radial outwards kick while the other half ends up as heat. 
        
        SMBH accretion in \simba includes two modes where cold rotationally supported gas accretes via a gravitational torque model \citep{Hopkins+Quataert:2011, DAA:2017a} while hot pressure supported gas follows the \citet{Bondi:1952} prescription. 
        In both of these modes accretion is suppressed by a radiative efficiency of 0.1.
        The torque based accretion is capped at 3 times the Eddington limit while the Bondi accretion mode strictly follows the Eddington limit.
        Additionally SMBHs are limited to grow no more than 0.1\% of their current mass in a single time step.
        The efficiency at which material is ejected in AGN winds is determined by the desired momentum input $\dot{P}_{\rm out} = 20L/c$ with $L = 0.1 \dot{M}_{BH} c^2$.
        For a $10^{9} M_\odot$ SMBH ejecting gas at maximum velocities this translates to an accreted mass to energy released conversion fraction of about 0.003 in the radiative mode or 0.03 in the jet mode.
        Additional information on the AGN feedback model can be found in \citet{dave:2019} and \citet{Christiansen:2020}. 
    
    \subsubsection{Different \simba Runs}

        Along with the 100 Mpc/h box ``flagship'' run, additional \simba simulations were run where the AGN feedback modes were turned off, one at a time.
        These simulations enable us to isolate the impact of each feedback mode. 
        \citet{Christiansen:2020} explored these feedback variant simulations and found that the AGN jet feedback in particular is vital in reproducing IGM properties at low redshift. 
        All of these additional simulations were run in a 50 Mpc/h box with $2\times512^3$ resolution elements.
        Apart from the variations in the AGN feedback, the remaining properties of these smaller box size simulations are the same as the flagship simulation.
    
        Among the extra runs they explored, we utilize one for comparison in this study. 
        The no-jet simulation removes the extra jet velocity boost from the AGN feedback so that the radiative feedback mode is the only AGN feedback present. 
        Since the X-ray feedback only occurs in conjunction with the jet mode it is also removed from the no-jet simulation. 
        \citet{Christiansen:2020} found negligible effects on the IGM when removing just the X-ray feedback, and
        we additionally found that removing X-ray feedback had a negligible effect on the \Lya~CDD.
        When removing the radiative feedback, \citet{Christiansen:2020} found little to no difference in the diffuse IGM at $z=0$, and we confirmed there is no substantial difference in the \Lya~CDD when removing AGN radiative feedback.
        As such, in this study we will focus on the no-jet variant and the fiducial 50 Mpc/h box \simba run for comparison.
        
    \subsection{IllustrisTNG}
    
        The TNG suite consists of magnetohydrodynamic cosmological simulations which vary in mass resolution, volume, and complexity of the physics included \citep{Pillepich:2018,Marinacci:2018,Naiman:2018,Springel:2018,Nelson:2018,Nelson:2019a,Pillepich:2019,Nelson:2019b}. 
        The simulations are performed with the AREPO code \citep{Springel:2010,Weinberger:2020} and gravitational interactions evolve via the TreePM algorithm \citep{Springel:2005}.
        Radiative cooling from hydrogen and helium is implemented using the network described in \citet{Katz:1996} and includes line cooling, free-free emission, and inverse Compton cooling.
        IllustrisTNG assumes ionization equilibrium and accounts for on-the-fly hydrogen column density shielding from the radiation background \citep{Rahmati:2013}. 
        Metals and metal-line cooling are included \citep{Vogelsberger:2012,Vogelsberger:2013} and star formation is implemented using the \citet{Springel:2003} subgrid model. 
        
        In this paper we use a small box run of TNG from the CAMELS project which uses the same sub-grid models as the original IllustrisTNG simulations \citep{camels:2021}. 
        We use this small box run instead of TNG100 since we compare to another small box run that removes the AGN kinetic feedback. 
        Since each run has the same box size, resolution and initial conditions, it makes for simpler comparisons. These runs will be explained further in section \ref{sec:diffTNGruns}. 
    
    \subsubsection{TNG AGN Feedback}
    
        The AGN feedback in TNG has a high Eddington ratio thermal mode and a low Eddington ratio kinetic mode. 
        In each mode the energy is directly deposited into the gas within the SMBH `feedback region'. The feedback region is a sphere around the SMBH with a size that scales with resolution $\propto m_{baryon}^{-1/3}$. The size of the feedback region is roughly constant within each simulation varying only slightly depending on the particles neighboring the SMBH \citep{Weinberger:2017, Pillepich:2018b}.
    
        The thermal mode deposits energy continuously as thermal energy. 
        The kinetic mode is significantly more efficient than the thermal mode, as the pulsed injection of energy in the kinetic mode heats up the gas in the feedback region to higher temperatures \citep{Weinberger:2017}.
        There is also a continually active radiative mode, which adds the SMBHs' radiation flux to the cosmic ionizing background.
        However, the effect of the radiative mode on the halo is limited to the brightest AGN and is fairly small \citep{Zinger:2020_feedback}.
    
        The transition to the kinetic feedback mode happens for SMBHs with $\eta < \chi$, where
    
        \begin{equation}
            \chi = \textnormal{min}\left[0.002 \left(\frac{M_{\rm BH}}{10^8 M_\odot}\right)^2,\ 0.1\right].
        \end{equation}
    
        \noindent Thus, typically BHs with masses greater than $M_{\rm BH} \sim 10^8 M_\odot$ produce kinetic feedback. 
        In this mode, energy is stored when the SMBH accretes gas until a minimum energy is reached ($E_\textnormal{inj,min} = 10 \sigma_{DM}^2 m_\textnormal{enc}$, where $\sigma_\textnormal{DM}^2$ is the one-dimensional dark matter velocity dispersion around the SMBH and $m_\textnormal{enc}$ is the gas mass in the feedback region).
        Once the threshold is reached, energy is injected in a random direction as a momentum kick to the gas within the feedback region. 
        
        SMBH accretion in the TNG uses the \citet{Bondi:1952} accretion prescription and is capped by the Eddington limit with a radiative efficiency of 0.2.
        The efficiency fraction (at which accreted mass is converted into energy for the thermal mode) is a constant $0.02$.
        In the kinetic mode the efficiency fraction is calculated as $\rho/(0.05\rho_{\rm SFthresh}$), where $\rho$ is the density of the gas around the SMBH and $\rho_{\rm SFthresh}$ is the star formation threshold density, but the efficiency fraction is capped at 0.2.
        See \citet{Weinberger:2017} and \citet{Zinger:2020_feedback} for more information on AGN feedback in TNG.
    
    \subsubsection{Different TNG Runs and CAMELS}\label{sec:diffTNGruns}
    
        For this study we explore small box TNG runs from the CAMELS project \citep{camels:2021}. 
        These simulations use the same sub-grid models as TNG and are essentially small box runs with a resolution comparable to the original TNG300-1 simulation. 
        Each CAMELS simulation has $256^3$ gas resolution elements in a periodic comoving volume with a side length of 25 Mpc/$h$. 
        The CAMELS project simulations are publicly available \citep{CAMELS-public}.
        
        We utilize two runs from the CAMELS project.
        The first is a publicly available simulation that employs the same physics as TNG but in a smaller box.
        The second is the same except it removes the AGN kinetic feedback mode and is not publicly available.
        Together these runs allow us to analyze the effect of turning off the strongest AGN feedback model in TNG as we do in the case of \simba. 
    
    \subsection{AGN Models in TNG vs.\ \simba}
    
        Of the different AGN feedback modes modeled, the kinetic mode in TNG and jet mode in \simba are the strongest feedback modes and thus are the most likely to affect the IGM (see Figure \ref{fig:Tprojs}). 
        While both modes are kinetic energy based and observationally motivated, they are vastly different in their implementation. The differences are:
    
        (1) The \simba jet mode hydrodynamically decouples, which allows the ejected gas to be deposited at some distance from the SMBH. By contrast, in TNG the kinetic energy of the feedback is deposited in the region immediately around the SMBH.
       
        A common struggle in cosmological simulations is producing outflows that can reach observationally motivated distances without over heating the gas but still produce galactic star formation rates that match observations.
        The low numerical resolution of cosmological simulations can hamper galactic outflows, and without decoupling the outflows can be mostly quenched in high-mass galaxies \citep[$M_h \sim 10^{12} M_\odot /h$,][]{DallaVecchia:2008}. 
        Decoupling in hydrodynamical simulations was first used for supernova feedback winds in \citet{Springel:2003}. 
        A momentum driven decoupled galactic outflow model has been found to successfully reproduce observed IGM carbon enrichment \citep{Oppenheimer:2006, Oppenheimer:2008}. 
        The \simba simulations included decoupling for the AGN jet feedback to mimic observed radio jets and deposit the energy of said jets well outside the galactic center \citep{dave:2019}. 
        
        (2) The TNG kinetic mode stores energy for discrete feedback events while the \simba AGN feedback is instantaneous. 
        In TNG the ejection speed in the kinetic mode is affected by the energy ejected into the surrounding medium which depends on the minimum energy required for an event to take place. 
        The wind speed at ejection depends on the energy ejected and the amount of mass being ejected, it is not explicitly set in the simulation. 
        Conversely, the jet speed in \simba is explicitly set based on the instantaneous SMBH mass and the Eddington ratio.
        
        (3) The jet feedback in \simba is collimated while the TNG kinetic feedback is isotropic.
        In \simba the ejection is bipolar occurring along the axis aligned with the angular momentum vector of the gas in the BH kernel. 
        In TNG the kinetic mode ejection occurs in a random direction that averages to isotropic over multiple events.
        
        (4) The \simba simulations use a two mode accretion method while the TNG simulations use a single Eddington limited Bondi accretion mode. As a result SMBHs in \simba can accrete more mass in a time step than those in TNG. However, \simba limits SMBHs to grow no more than 0.1\% of their current mass in any given time step.
        Additionally the radiative efficiency fraction is different between \simba and TNG with \simba having a lower radiative efficiency fraction (0.1 in \simba and 0.2 in TNG).
        
        (5) The \simba simulations heat AGN jets to the virial temperature of the host halo. The kinetic mode in TNG also heats up the gas ejected, occasionally beyond the virial temperature, but the temperature is not explicitly set as opposed to \simba.
    
       In addition to the differences between the two feedback models, the UVB used in each simulation is also different. 
        The UVB model also has an effect on the \Lya~forest statistics so it must be considered here.
        \simba uses a modified version of the \citet{Haardt:2012} UVB implemented using the GRACKLE framework \citep{Grackle:2017} while TNG uses the \citet{Faucher-Giguere:2009} UVB which has a larger UV photon density at $z = 0.1$.
    
  \subsection{Generating Column Densities}\label{sec:gen_CDDF}
    
     In this study we generate column density sightlines from the simulations using the publicly available \textit{fake-spectra}\footnote{\url{https://github.com/sbird/fake_spectra}} code outlined in \citet{Bird:2015, Bird:2017} and with MPI support from \citet{Qezlou:2022}. The code generates and analyzes mock spectra from simulation snapshots and is fast, parallel, and native-SPH. It is written in C++ and Python 3 with the user interface being Python-based.
     
     The neutral hydrogen column densities ($N_{\rm HI}$) from our generated sightlines are calculated for each pixel in units of neutral hydrogen atoms (HI) cm$^{-2}$. 
     Column densities are computed by interpolating the neutral hydrogen mass in each gas element to the sightline using an SPH (smoothed particle hydrodynamics) kernel. The method used is based on the type of simulation; for the CAMELS TNG simulations a tophat (or uniform) kernel is used while for \simba a cubic spline kernel is used.
     
     The particles in 525 kpc/h slices are amalgamated into absorbers which are then used to calculate column densities for the CDD. 
     We find that defining the absorber size around 500 kpc/h results in a well converged CDD that simultaneously covers the smallest CD absorbers ($N_{\rm HI} \lesssim 10^{13}$ cm$^{-2}$) without slicing larger CD absorbers.
     Changing this value by a factor of 2 has a $\lesssim$ 10\% effect on the CDD at column density values smaller than $10^{12.5}$ cm$^{-2}$. 
     The CDD at column density values larger than $10^{14}$ cm$^{-2}$ is insensitive to the choice of absorber size so long as that choice is not overly large  (i.e. on the order of Mpc). We note that \citet{Gurvich2017} demonstrated that the results from this column density calculation method are strikingly similar to the results from a full Voigt profile fit analysis.
     
     We generate 5,000 sightlines randomly placed in each simulation box. 
     We find this number of sightlines to be sufficient to avoid variations due to sampling. 
     We note that at column densities of $N_{\rm HI} > 10^{15} \textnormal{cm}^{-2}$ variations in the CDD will increase due to the rarity of those absorbers, however the observational error bars at these densities are also significantly larger.
     
\section{Results} \label{sec:res}

    Figures \ref{fig:Tprojs} and \ref{fig:CDprojs} show a mass weighted temperature projection and a \Lya~forest column density projection respectively for the different simulations we study. 
    These plots help illustrate the differences between the \simba and TNG simulations when disabling the strongest AGN feedback mode. 
    These projections are over a slice with thickness corresponding to the size of an absorber as defined in this study ($525$ kpc/h). 
    
    From the temperature projections it is clear that both the \simba and TNG AGN feedback models have an effect on the temperature distribution in the simulations. 
    However the \simba jet feedback propagates much further through the simulation box than the kinetic mode in TNG. 
    
    In Figure \ref{fig:CDprojs}, the column density projections demonstrate a clear difference between the \simba jet and no-jet runs but minimal difference is seen in the TNG simulations. 
    From these projections alone we should expect to see a clear difference in the CDD for the different \simba runs but minimal difference for the TNG runs.

    The column density distribution function ($f(N_{\rm HI})$) is defined as 
     
     \begin{equation}
     f(N_{\rm HI}) = \frac{d^2N}{d\log (N_{\rm HI}) dz} = \frac{F(N_{\rm HI})}{\Delta N_{\rm HI}} \Delta z  
     \end{equation}
     
     \noindent where $F(N_{\rm HI})$ is the fraction of absorbers with column densities in the range [$N_{\rm HI}$, $N_{\rm HI}+\Delta N_{\rm HI}$], and $\Delta z$ is the redshift distance of the sightline. 
     The CDD describes the number of absorbers within a logarithmic column density bin width and redshift distance.

    \begin{figure*}
        \centering
        \includegraphics[width=0.55\linewidth]{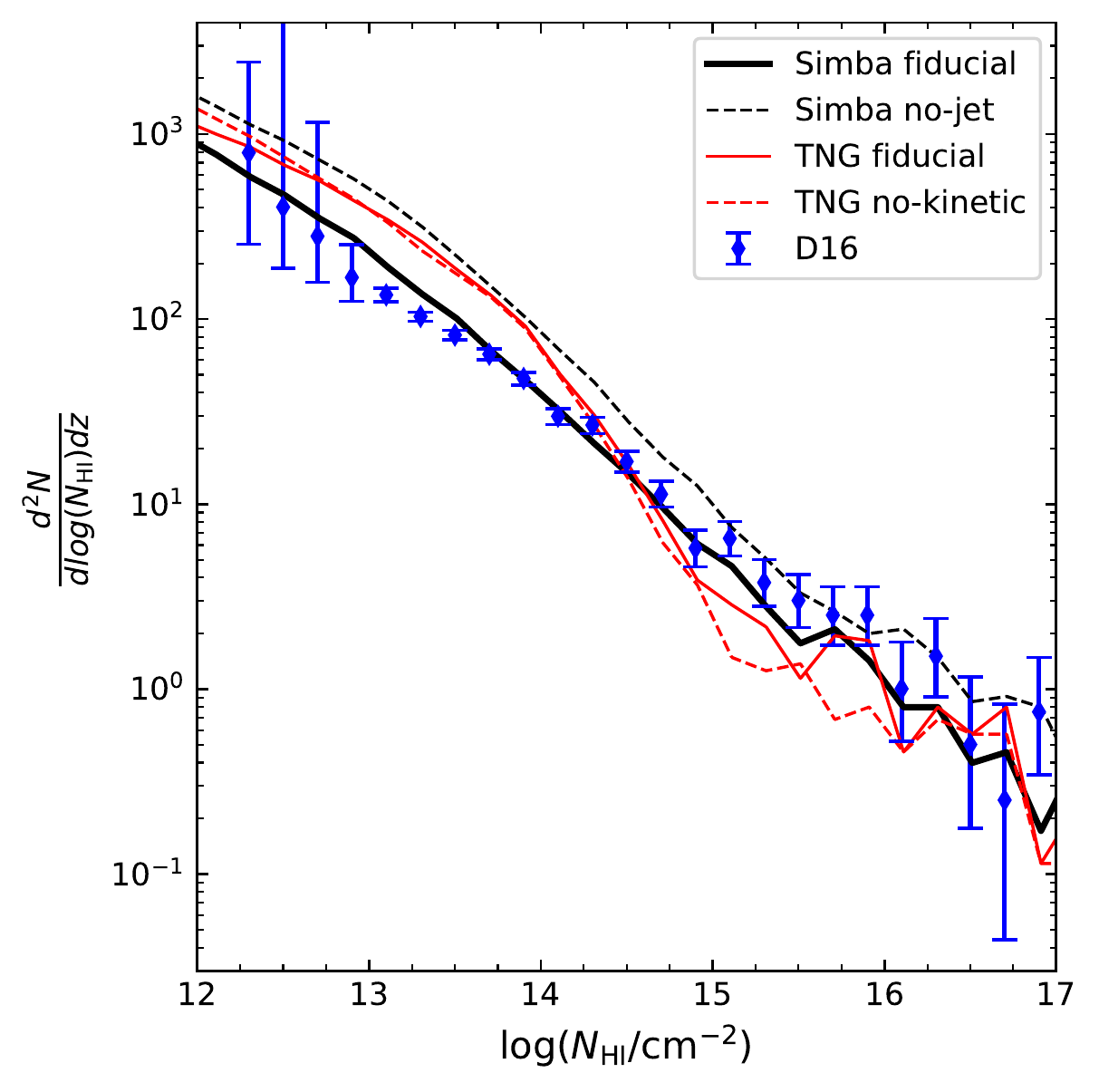}
        \caption{The $z=0.1$ CDDs for the \simba runs (black lines) and the IllustrisTNG runs (red lines), compared to the D16 observational data (blue points). The solid lines are the fiducial runs while the dashed lines are the no-jet/kinetic runs. The observational data is from Table 6 of D16.}
        \label{fig:CDDF} 
    \end{figure*}

    The $z=0.1$ CDDs produced for the various \simba and TNG simulations are presented in Figure \ref{fig:CDDF} along with the D16 observational data. The data from the D16 catalog covers a redshift range of $z=0 - 0.47$ with 65\% of the absorbers coming from the range $z=0-0.2$ and the median redshift for absorbers is $z=0.14$. We have explored producing CDDs with contributions from the different redshift bins seen in the D16 catalog and found that this work's main findings remain unchanged. The simulated CDD for both \simba and TNG are within 1$\sigma$ (relative to the  D16 data error bars) for redshifts $0<z<0.2$ and within 2$\sigma$ for $0.2<z<0.4$.
    
    Looking at Figure \ref{fig:CDDF}, the main differences between the \simba and TNG results are the slope of the CDD in the range $10^{14} < N_{\rm HI} < 10^{15}$ cm$^{-2}$ and the normalization at $N_{\rm HI} < 10^{14}$ cm$^{-2}$. TNG tends to over-predict the number of low CD absorbers ($N_{\rm HI} \lesssim 10^{14}$ cm$^{-2}$) as compared to \simba and D16.
    The $N_{\rm HI} \gtrsim 15$ cm$^{-2}$ normalization of fiducial TNG appears to match the observational data approximately, however larger box runs of TNG (where $N_{\rm HI} > 10^{15}$ statistics are more robust) show that those CD absorbers are actually slightly under-predicted \citep{Burkhart_2022}. 
    Changing the strength of the UVB model could reduce the disparity of the normalization offset at low CDs but would increase the disparity at high CDs (and vice versa), and the UVB cannot affect the slope of the CDD which is the main cause for a poor fit to the D16 data.
    The \simba CDD has an overall shallower slope and shows a remarkable match to the observations.

    \begin{figure}
        \centering
        \includegraphics[width=0.9\linewidth,trim={0 0 0 0},clip]{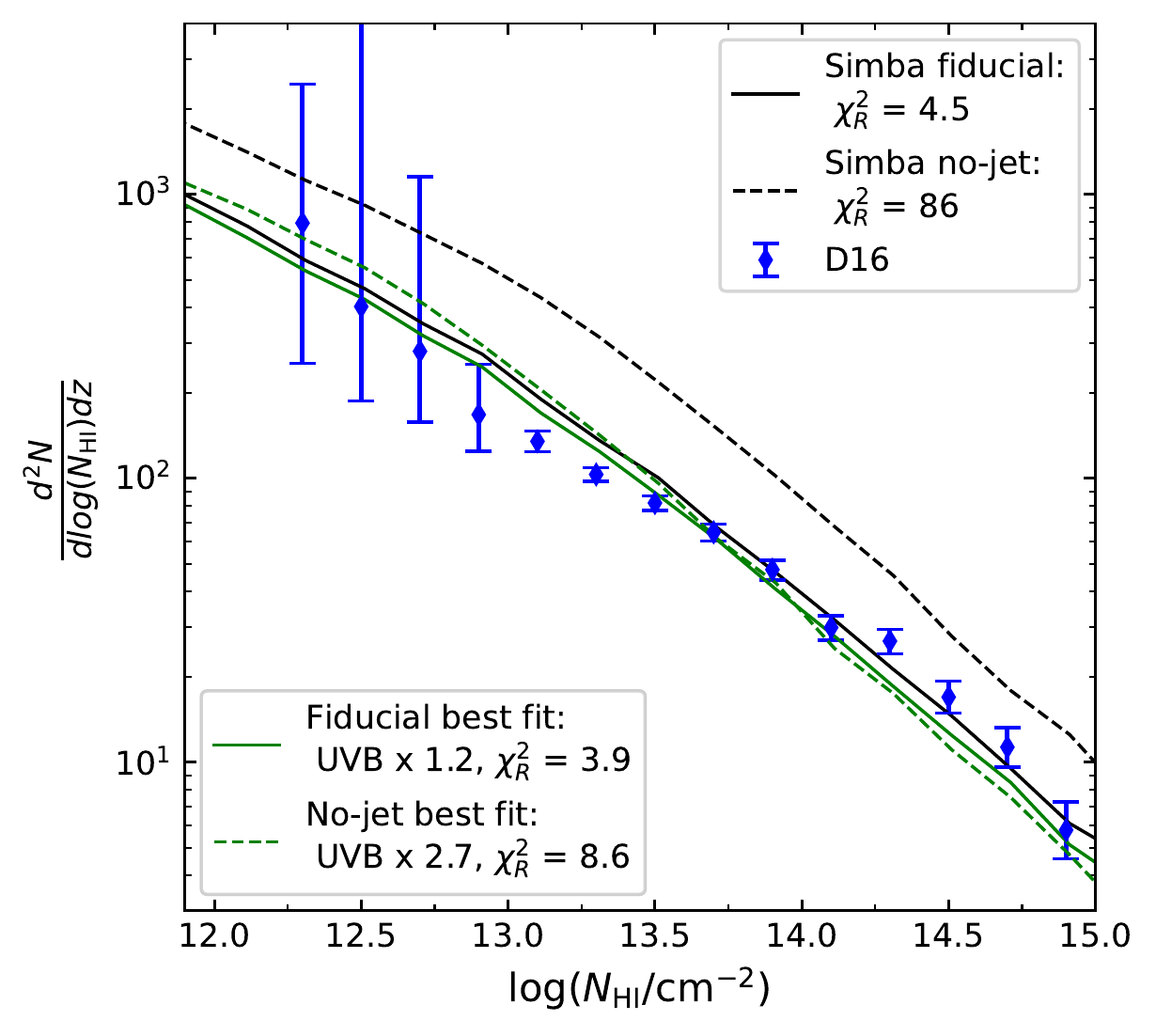}
        \caption{The \simba $z=0.1$ CDDs corresponding to the raw simulation results (black lines) and when correcting the UVB post-processing to find the best-fit (green lines). 
        Results are shown for both the fiducial (solid lines) and no-jet runs (dashed lines). 
        The $\chi^2_R$ values and the UVB correction factors (relative to the \citet{Haardt:2012} UVB) for the fits are in the legends.}
        \label{fig:UVB} 
    \end{figure}
    
   In order to investigate the effect of the AGN jet mode we also over-plot the CDD of the no-jet \simba simulation (black dashed line).  
   The no-jet \simba simulation shows dramatic differences from the fiducial \simba run (i.e.\ with jets) at all column densities.
   The main effect here is a re-normalization of the CDD implying much more neutral hydrogen is present in the IGM but the AGN jet feedback in \simba also has a secondary effect on the CDD slope.

\section{Discussion} \label{sec:dis}

\subsection{Goodness of Fit and UVB Corrections}
    We have found that the fiducial \simba run is an excellent match to the CDD from D16.  
    By comparison with the no-jet \simba run, we find that AGN jet feedback as implemented in \simba is a viable solution for resolving the discrepancy between the observed and simulated low redshift \Lya~forest \citep[supporting results from][]{Christiansen:2020}. 
    Since the UVB also sets the amplitude of the CDD, we conduct a least squares fit to find the UVB correction factor required for the best fit of the \simba CDDs to the D16 observational data.
    For our fitting procedure, we assume Gaussian distributed random variables.
    Additionally, by conducting this fit on both the fiducial and no-jet \simba runs we can further disentangle the effects of the UVB vs.\ AGN jet feedback on the low-$z$ CDD.
    We conduct this fit within a CD range of $N_{\rm HI} = 10^{12}$ to $10^{15}$ cm$^{-2}$ as this is where the simulation data is most robust and we find that well motivated variations in this range do not qualitatively affect the fit results (see Appendix \ref{apendix:fitrange} for additional details).
    
    We determine the \simba CDD resulting from a different UVB model using a post-processing correction method.
    We follow a similar procedure as the one outlined in \citet{Kollmeier:2014} which uses the approximation that $N_{\rm HI} \propto 1/\Gamma_{\rm HI}$ where $\Gamma_{\rm HI}$ is the hydrogen photoionization rate.
    This approach works since the low redshift \Lya~forest can be well approximated as an optically thin region in photoionization equilibrium.
    This method breaks down when absorbers are no longer optically thin but is well converged for CDs up to at least $N_{\rm HI} \sim 10^{15}$ cm$^{-2}$ (with the limiting factor being the simulation box size) and can be applied in post-processing.
    
    Looking at Figure \ref{fig:UVB}, the best-fit UVB correction factor for the no-jet run is 2.7 times stronger than the \citet{Haardt:2012} model and the correction factor for the fiducial run is 1.2 times stronger. 
    When calculating our reduced $\chi^2$ ($\chi^2_R$), the number of degrees of freedom is the number of observational points being fit and the only variable parameter is the UVB correction factor. For the no-jet run $\chi^2_R =$ 8.6 while for the fiducial run $\chi^2_R =$ 3.9. 
    The no-jet $\chi^2_R$ is $\sim$2.2 times larger than that of the fiducial run and while both fits fail to produce a $\chi^2_R$ value close to 1 these results show that the fit with AGN jets included is quantitatively better than without jets. 
    These result exemplify the importance of the slope change to the CDD when including AGN jets, especially in a range where the observational data is most robust. 
    
    While both fits result in statistically bad $\chi^2_R$ values, only the fiducial best fit is preferred over both of the $\chi^2_R$ values of the raw fiducial and no-jet simulation results (i.e.\ when the strength of the UVB is \textit{not} allowed to vary). 
    The raw \simba fits produces $\chi^2_R =$ 4.5 for the fiducial simulation, and $\chi^2_R =$ 86 for the no-jet simulation. 
    While adding jets to the simulation produces a $\sim 19$ times lower $\chi^2_R$ value, varying the UVB further improves the $\chi^2_R$ value. 
    These results emphasize how important it is to consider the UVB in conjunction with the AGN feedback effects.
    Additionally, the no-jet best fit when allowing the UVB to vary produces a $\chi^2_R$ value nearly twice as large as the raw \simba fiducial fit.
    Even when the UVB is allowed to vary, the no-jet \simba results can not produce a better fit to the data than the fiducial \simba results.

    Recent studies have found hydrogen photoionizing values at $z=0.1$ that are $\sim$ 1.77, 1.78, 2.56, and 1.74 \citep[for][respectively]{Gaikwad:2017, KhaireUVB:2019, Puchwein:2019, FG:2020} times stronger than the \citet{Haardt:2012} values. 
    However it has been shown that this factor could go as high as $\sim 5$ when allowing the escape fraction of HI ionizing photons from galaxies to vary \citep{Khaire+Srianand:2015}.
    The best fit UVB of the fiducial simulation is 1.2 times stronger than \citet{Haardt:2012} and 1.2 times weaker than \citet{Faucher-Giguere:2009}. By comparison, the no-jet fit requires a UVB 2.7 times stronger than \citet{Haardt:2012}, which is 1.8 times stronger than \citet{Faucher-Giguere:2009} and slightly larger than \citet{Puchwein:2019} (these three models being some of the most commonly used UVB models in cosmological simulations).

\subsection{\simba vs.\ TNG: the UVB and AGN 
Feedback}

    As previously mentioned, the different UVB models (in addition to the AGN feedback models) utilized in TNG vs.\ \simba create a discrepancy in their resulting CDDs.
    The TNG UVB is $\sim 1.5$ times stronger than the \simba UVB at $z=0.1$. 
    Correcting the \simba results to the stronger UVB results in an overall normalization shift of the CDD downwards.
    Since changing the UVB largely means changing the normalization of the CDD, we cannot attribute the differences in the CDD slopes to the UVB models utilized.
    Additionally, the correction makes the resemblance between the \simba no-jet and the fiducial TNG results more apparent.
    
    We highlight that although kinetic feedback is included in both \simba and TNG, there remains a dramatic difference in their CDD shapes.
    This indicates that the implementation of the jet feedback sub-grid model is important to the resulting \Lya~forest statistics \citep[as seen in][comparing Illustris to IllustrisTNG]{Burkhart_2022}. 
    When removing the jet feedback from \simba there is a clear steepening of the CDD slope at $N_{\rm HI} < 10^{14.5}$ cm$^{-2}$, but the change is not enough to explain the large difference from the TNG slope around those CD values. 
    While it remains difficult to determine the full extent of this difference at higher column densities ($N_{\rm HI} > 10^{15}$ cm$^{-2}$), it is likely that additional factors apart from AGN jet feedback are also influencing the value of the CDD slope.

    We argue that the main reason \simba jet feedback affects the \Lya~forest while the TNG kinetic mode does not is the distance the AGN jets can travel.
    The decoupled AGN jets in \simba can travel up to $\sim 10$ kpc before they begin to deposit their energy \citep{dave:2019}, which means these jets can bypass much of the ISM. 
    Since the majority of the AGN jet energy can still be present by the time it reaches the CGM, it has a much higher likelihood of being able to propagate to the diffuse IGM.
    In TNG, \citet{Zinger:2020_feedback} found that galaxies around $10^{10.5} M_\odot$ can produce kinetic feedback that heats up the ejected gas to temperatures beyond $T_{\rm vir}$ and can even completely remove that gas from the galaxy.
    Despite this, the results from Figure \ref{fig:Tprojs} clearly show that energy from TNG kinetic feedback cannot propagate as far into the IGM as \simba jet feedback. Instead, TNG kinetic feedback largely affects the host galaxy and the more immediate surroundings. 
    Additionally, looking at Figure \ref{fig:CDprojs} we see minimal effect on the TNG column densities when adding/removing kinetic feedback, but we see a dramatic effect for \simba when jet feedback is toggled.
    
    While the column densities plotted in Figure \ref{fig:CDDF} are all traditionally considered part of the \Lya~forest, and assumed to be found largely in the IGM, the higher column densities ($N_{\rm HI} > 10^{13.5}$ cm$^{-2}$) tend to exist closer to and within halos/galaxies \citep{Bouma:2021} and the circumgalatic medium (CGM) \citep[as in COS-Halos][]{Werk:2014,Prochaska:2017}.
    Therefore, the slope change at $N_{\rm HI} \sim 10^{14.5}$ cm$^{-2}$ when removing the jet feedback in \simba implies that with decoupling some jets may be capable of completely bypassing the material located within the host halo.
    Since the slope change is subtle the fraction of jets able to bypass the entirety of the halo would be small.
    
    Looking at the maximum AGN wind speeds further motivates the argument that \simba jets reach the diffuse IGM while TNG kinetic feedback does not. 
    \simba AGN outflows can reach maximum velocities of $\sim 8000$ km/s (the outflow velocity in the radiative mode can reach up to $\sim 1000$ km/s plus the $7000$ km/s boost in jet mode) which is deposited as far as $\sim 10$ kpc away from the point of ejection \citep{dave:2019}. 
    Energy loss due to gravity results in a maximum velocity of $\sim 7000$ km/s at the point jets recouple. 
    The wind velocity at injection in TNG is not explicitly set but instead depends on the energy ejected and the amount of mass to which the momentum kick applies. 
    \citet{Nelson:2019a} found in post-processing analysis that TNG AGN winds reach maximum velocities of $\gtrsim 12,000$ km/s at time of injection (much higher than in \simba) but this reduces to $\sim ~ 3,000$ km/s at a distance of 10 kpc from the injection site.
    This is less than half the speed jets in \simba exhibit at a similar distance.
    
    Another factor that could be reducing the distance that the TNG AGN feedback effects propagate to is the direction of ejection.
    TNG AGN kinetic feedback is ejected in a random direction which averages to isotropic ejection over many events. 
    Random directions that are more parallel to the galactic disk may reduce the overall distance reached due to quenching of AGN winds as they travel through the disk. 
    In contrast, \simba jets are highly collimated along the angular momentum vector of gas within the black hole kernel. 
    While this direction can change slightly over time, recurrent ejection of gas along a preferred direction can help maximize the long-range impact of AGN feedback.
    
    In summary, from our comparison between the \simba fiducial and \simba no-jet CDDs, we find that jet feedback in \simba lowers the CDD at all column densities, even the higher column densities associated with the CGM in halos \citep{Werk:2013}.
    Therefore, the AGN jet feedback in \simba can re-couple to its surroundings close enough to the injection site to affect the CGM while still being able to reach as far as the diffuse IGM.
    Removing the AGN kinetic feedback mode from TNG has a negligible effect on the CDD implying that TNG jets are unable to reach/affect most if not all of the absorbers that make up the low-$z$ \Lya~forest.
    
    An important note is that the highest CDs ($N_{\rm HI} > 10^{15}$ cm$^{-2}$) should be subject to more scrutiny as these absorbers are less common and more likely to be affected by simulation box size and initial conditions.
    In the appendix we show these values to be converged between the original and small box simulations within 1$\sigma$ of the observational data.
    The TNG no-kinetic feedback run at the higher CD ranges also appears largely converged with the original TNG100 simulation.
    
    As it stands in this work, the TNG AGN kinetic feedback appears to have no effect on the low redshift \Lya~forest (see Figure 3).
    To truly disentangle AGN feedback effects on the neutral hydrogen in the TNG simulation one would need to study even higher CDs such as Lyman limit systems ($10^{17} < N_{\rm HI} < 10^{20}$ cm$^{-2}$) and damped \Lya~absorbers ($N_{\rm HI} > 10^{20}$ cm$^{-2}$) in the fiducial vs.\ no-kinetic feedback TNG runs.
    However, obtaining a robust statistical sample of these types of absorbers requires a larger simulation box size.
    
    The \simba CDD demonstrates that the mismatch between the observed and simulated \Lya~forest can be solved by a combination of heating from AGN jet feedback and a slightly harder UVB at z=0.1 \citep[in relation to][]{Haardt:2012}.
   As indicated by the results in Figure \ref{fig:UVB}, with AGN jet feedback there is the potential of overheating the IGM when using stronger  UVB models \citep[with $\gtrsim 1.75$ times stronger UVBs in][]{Gaikwad:2017,KhaireUVB:2019, Puchwein:2019, FG:2020}. Since the forest at z=0.1 is sensitive to both AGN heating and the UVB, additional constraints of the low redshift UVB using HI and H-$\alpha$ are of great value \citep{Adams:2011, Fumagalli:2017}. 
    
    By design, AGN models in  both the \simba and TNG simulations are in good agreement with the observed stellar mass function and specific star formation rate to stellar mass relation at lower redshifts \citep{Romeel:2020}. 
    This emphasizes that an AGN jet feedback model, when implemented carefully, can reproduce not only observed intergalactic properties but several observed galactic properties as well.
    
    Finally, it is important to acknowledge that while an AGN feedback model may be able to produce well-converged galactic statistics it is also necessary to consider the intracluster and intragroup statistics. 
    Several studies have explored AGN feedback in the context of galaxy groups and clusters and have found the \simba's AGN jet feedback plays a particularly unique role in the determination of the X-ray statistics \citep{Robson:2020, Robson:2021, Yang:2022}.
    Notably the fiducial \simba simulations are in good agreement with the observed hot baryon fractions as a function of halo mass. 
    However, other group/cluster statistics struggle to match what is observed such as the entropy profiles \citep{Oppenheimer:2021}.
    Recurrent energetic ejections via AGN jets can have a catastrophic effect on galaxy group statistics.
    Recent studies have emphasized the importance of considering these statistics when constraining future AGN feedback models \citep{Oppenheimer:2021,Lovisari:2021}.

\section{Conclusions} \label{sec:con}

    We analyze the $z=0.1$ \Lya~forest column density distribution (CDD) for the \simba and IllustrisTNG simulations to explore the effect of the different AGN feedback subgrid models. 
    Additionally, we analyze variations of these simulations that remove the strongest AGN feedback modes to determine the feedback's effect on the CDD and thus the IGM.
    
    We confirm findings by \citet{Christiansen:2020} that AGN jet feedback is potentially vital to include in future cosmological hydrodynamic simulations as a partial solution to the low redshift \Lya~forest discrepancy between simulations and observations. 
    We also support findings from \citet{Burkhart_2022} that the precise implementation of the AGN feedback sub-grid model is a point worth considerable attention.
    Despite both \simba and TNG implementing observationally motivated models for AGN feedback, the \simba results match the observed CDD remarkably while TNG struggles with both slope and normalization offsets from the observed data. 
    
    Our main conclusions are as follows:
    
    \begin{itemize}
    
    \item The fiducial \simba cosmological simulations, which employ the \citet{Haardt:2012} UVB model and a three-mode AGN feedback model with radiative, jet, and X-ray modes, provides a remarkable match to observational data from D16. The low redshift ($z=0.1$) \Lya~forest CDD in \simba matches HST COS data from D16. 
    
    \item We conclude that AGN jets in \simba are able to inject heat and energy far away from the host halos and into the diffuse IGM. By z=0.1 these jets produce a better match to the low-$z$ \Lya~CDD than the same simulation without jets at all CDs. This holds true even when a best-fit UVB correction is conducted on the no-jet CDD. Future explorations regarding the constraint of the low-z \Lya~forest should consider AGN jet feedback as a potential factor contributing to IGM heating.
    
    \item We argue that the long-range AGN jets in \simba, which result from many variables in the AGN feedback model such as the decoupling, temperature and velocity of  the ejected material, and collimation of the jets, allow the feedback effects to reach the diffuse IGM and affect the \Lya~forest.
    The TNG implementation of kinetic feedback results in ejections that are largely confined to the host halo and thus unable to affect the \Lya~forest CDD.
    
    \item We emphasize that AGN feedback models should be built and implemented with the UVB model utilized in mind.
    Although AGN jet feedback can affect the \Lya~forest in ways the UVB cannot (i.e.\ changing the slope of the CDD) these mechanisms have degenerate effects on the forest CDD. 
    AGN jet feedback that is too strong can risk overheating the IGM when combined with stronger UVB models. However, a slightly stronger UVB model than \citet{Haardt:2012} was required in addition to AGN jets to further improve the fit between HST COS data and the \simba simulation low-$z$ \Lya~forest.
    \end{itemize}
    
    Since the precise implementation of jet feedback can have a dramatic effect on the low redshift \Lya~forest statistics, exploring different subgrid models will be vital in constraining AGN feedback as a whole.
    Despite the AGN feedback having a clear effect on the CDD slope at $z=0.1$, we were not able to confidently infer the full reason for the slope difference between the \simba and TNG CDD.
    This task is difficult due to the large quantity of factors that can affect the neutral hydrogen distribution in the IGM.
    Given our results, it is clear the slope difference comes in part from the inclusion of jets in the \simba simulations, but jets do not fully explain the \simba vs.\ TNG CDD slope difference.
    Larger box runs (or additional runs with varying initial conditions) of the simulations where AGN jet feedback is turned off are likely necessary to fully converge the CDD at the highest column densities and determine the full extent of the \simba slope changes.
    These runs could also reveal any effects missed when removing the TNG kinetic feedback mode (e.g. at higher CDs ($N_{\rm HI} > 10^{15}$ cm$^{-2}$) where box-size becomes important).
    
    It will be important to explore the variations of specific simulations' sub-grid models in addition to comparing sub-grid models as a whole (e.g.\ investigate variations in the strength of TNG's AGN feedback rather than comparing TNG to \simba). 
    It is possible that the difference in the CDD slope between IllustrisTNG and \simba could further be explained by these variables.
    The CAMELS project, a collection of simulations varying astrophysical parameters within the framework of different simulation suites  \citep{camels:2021}, is an excellent starting point for exploring the subtleties of different AGN feedback sub-grid models.
    Exploration of the AGN feedback parameters is revealing an interesting interplay between AGN and stellar feedback and their effect on the low redshift \Lya~forest (Tillman et al. in prep).
    The exploration of AGN feedback parameters in CAMELS makes it a powerful tool for disentangling AGN feedback's potential role in converging the simulated and observed low redshift \Lya~forest.

\begin{acknowledgments}
MTT thanks the Simons Foundation and the Center for Computation Astrophysics at the Flatiron Institute for providing the computational resources used for this analysis. 
MTT thanks Mahdi Qezlou for their assistance in navigating the \textit{fake-spectra} package and for helpful conversations regarding proper usage and trouble shooting. 
BB is grateful for generous support by the David and Lucile Packard Foundation and Alfred P. Sloan Foundation.  This work is supported by NASA Astrophysics Theory Program grant number 80NSSC22K0823. 
DAA acknowledges support by NSF grants AST-2009687 and AST-2108944, CXO grant TM2-23006X, and Simons Foundation award CCA-1018464. 
GLB acknowledges support from the NSF (AST-2108470, XSEDE grant MCA06N030), NASA TCAN award 80NSSC21K1053, and the Simons Foundation.
\end{acknowledgments}

\appendix

\section{Resolution and Box Size}

\begin{figure}[h]
        \centering
        \includegraphics[width=0.6\linewidth,trim={0 0 4.9in 0},clip]{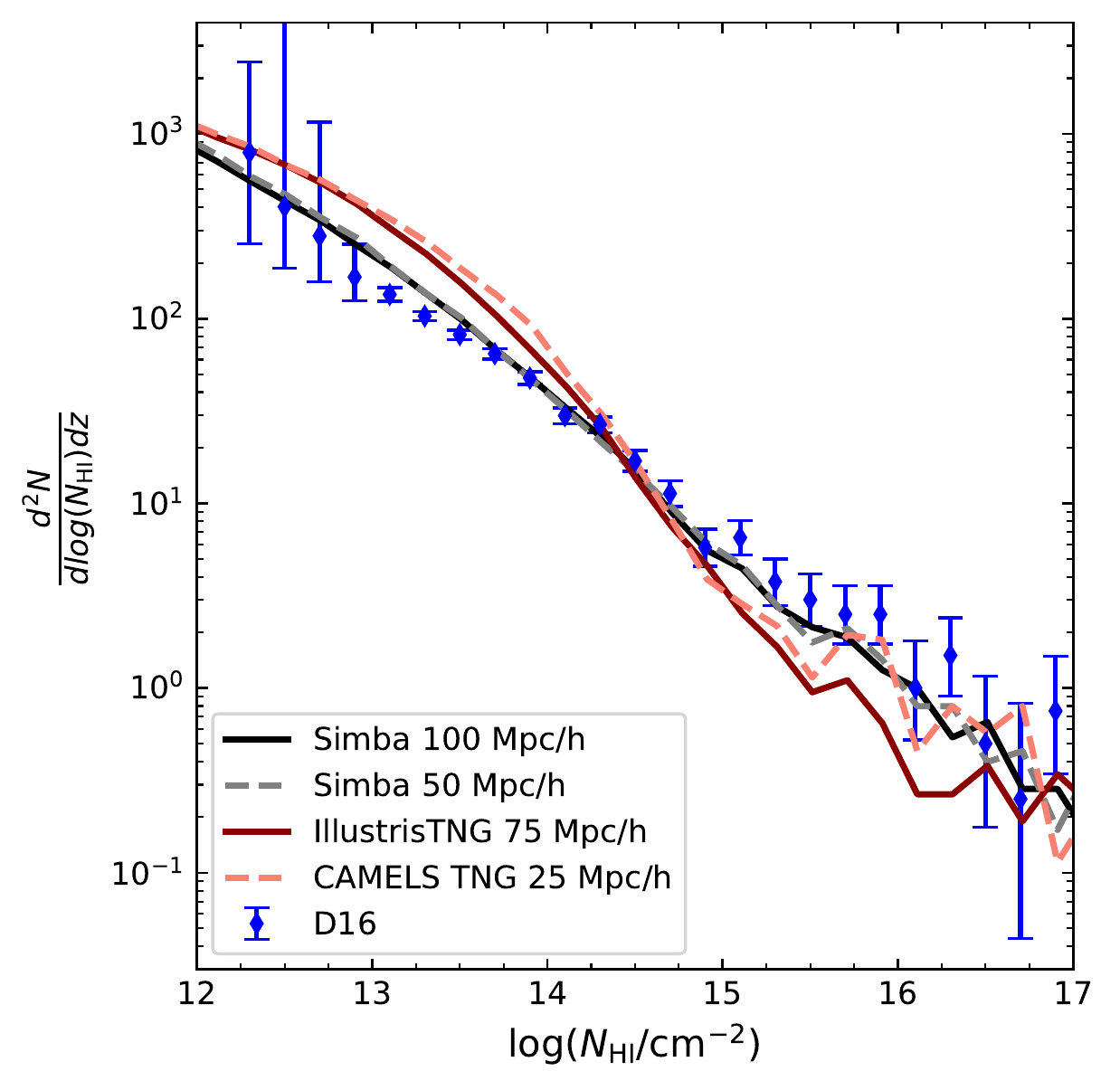}
        \caption{The $z=0.1$ CDD for the original \simba and TNG runs vs.\ the small box runs studied in this work. The results are converged exhibiting that the box size and resolution of these runs are converged for the \Lya~statistics presented within this study.}
        \label{fig:boxsize} 
    \end{figure}

    Figure \ref{fig:boxsize} shows the CDD for the small-box \simba and TNG runs explored herein vs.\ the original simulation runs.
    The small-box runs appear largely converged to the results of the full-box runs. 
    The \Lya~forest in TNG was studied in \citet{Burkhart_2022} and they found the CDD results to be converged for box sizes from 50 Mpc/h to 300 Mpc/h side lengths (at least up to $N_{\rm HI} \sim 10^{15}$ cm$^{-2}$). 
    Figure \ref{fig:boxsize} confirms the convergence of the \simba original 100 Mpc/h box size run and the small-box 50 Mpc/h run. The CAMELS TNG 25 Mpc/h box size run is converged with the original TNG100 run within 1$\sigma$ of the observational error bars. 
    The resolution is an important factor in the convergence of the \Lya~CDD with poor resolution leading to overall normalization shifts in the CDD that get worse with higher $N_{\rm HI}$ \citep{Burkhart_2022}. 
    The mass resolutions in this study are sufficient as the normalization effect does not arise between the different runs.
    
\section{Fitting Range}\label{apendix:fitrange}

For the fitting procedure conducted and discussed in Section \ref{sec:dis} we explored several well motivated variations in the CD range originally fit.
Note that in all the fits the D16 data point at $N_{\rm HI} = 10^{12.1}$ cm$^{-2}$ is thrown out due to the absent lower bound.
The resulting $\chi^2_R$ values from these range variations are presented in Table \ref{table}. Neglecting observations below CDs of $10^{13}$ cm$^{-2}$ was explored due to the fact that observational error bars are significantly higher in that range. 
Neglecting observations above CDs of $10^{14.75}$ cm$^{-2}$ was explored since the box size of the simulation can affect the post-processing UVB correction method. 
Due to the nature of the UVB correction method utilized and the fact that smaller box size simulations have overall less high column density absorbers ($\gtrsim 10^{15}$ cm$^{-2}$) the UVB correction can break down at the highest CDs given an insufficient box size. 
For the \simba simulation box size explored herein the approximate UVB correction is reliable up to (at least) CDs of $\sim10^{14.75}$ cm${-2}$.

For each of the fitting ranges explored, the results of this work remain qualitatively unchanged. The $\chi^2_R$ values vary for each range but the overall ratio between the fiducial (jet) and no-jet fits remain approximately the same.
Additionally, the fiducial fits, for both when the UVB is allowed to vary and when it is not, remain quantitatively better than the no-jet best-fits for all ranges explored.
Finally, the UVB correction factors do not change when the fitting range is allowed to vary (at least to 2 significant digits).
Therefore we conclude that reasonable variations in the fitting range do not affect the qualitative results of the fits.

\begin{table}[h]
\caption{The fitting ranges explore within, the corresponding $\chi^2_R$ values of the UVB best-fits, and the ratio of those values. The last column shows the $\chi^2_R$ values for the fiducial \simba results when the UVB is \textit{not} allowed to vary.}
\centering
\begin{tabular}{c|ccc|c}
CD Range Fit (log{[}$N_{\rm HI}$/$cm^{-2}${]})    &    No-jet best-fit $\chi^2_R$    &    Jet best-fit $\chi^2_R$    &    No-jet $\chi^2_R$/Jet $\chi^2_R$ & Fiducial $\chi^2_R$ \\ \hline \hline
12 - 15                              & 8.6             & 3.9          & 2.2 & 4.5                        \\
13 - 15                              & 11.9            & 5.5          & 2.2  &   6.0                     \\
12 - 14.75                           & 9.2             & 4.2          & 2.2   & 4.9                       \\
13 - 14.75                           & 13.3            & 6.1          & 2.2  & 6.7                       
\end{tabular}\label{table}
\end{table}

\bibliography{ms}{}
\bibliographystyle{aasjournal}
 
\end{document}